\documentstyle[12pt]{article} 
\advance\hoffset by -7mm  
\makeatletter
\@addtoreset{equation}{section}
\makeatother

\renewcommand{\title}[1]{\null\vspace{25mm}

\noindent{\Large{\bf #1}}\vspace{10mm}

\noindent {\large By }}
\newcommand{\authors}[1]{\noindent{\large #1}\vspace{3mm}

}
\newcommand{\address}[1]{\noindent #1\vspace{5mm}

}
\renewcommand{\abstract}[1]{\vspace{19mm}

\noindent{\small{\em Abstract.} #1}\vspace{2mm}

} 

\begin{document}

\title{Clock Synchronisation in Inertial Frames and Gravitational 
Fields}
\authors{F. Goy\footnote{Financial support of the Swiss National Science 
Foundation and the Swiss Academy of Engineering Sciences.}}

\address{Dipartimento di Fisica,
Universit\^^ {a} di Bari  \\
Via G. Amendola, 173 \\
I-70126  Bari, Italy \\ 
E-mail:  goy@axpba1.ba.infn.it}
 
\abstract{
The special relativistic test theory of Mansouri and Sexl is sketched. 
Theories based on different 
clock synchronisations are found to be equivalent to special relativity, as
regards experimental results. The 
conventionality of clock synchronisation is shown not to hold, by means of an
example, in a simple accelerated system and through the principle of 
equivalence in gravitational fields, especially when the metric is not static. 
Experimental implications on very precise clock synchronisation on earth are 
discussed.}

\section{The special relativistic test theory of Mansouri-Sexl}

As early as 1911, Frank and Rothe \cite{frro:11a} showed that 
the relativity principle alone 
implies the existence of a constant velocity in all inertial frames. If we 
assume that this velocity is the velocity of light, we see that the constancy
of the velocity of light is a consequence of the relativity principle.
Following Mansouri and Sexl \cite{mase:77a}, we want to construct a set of 
theories, which enable 
to express departures from the relativity principle. This set of theories 
consists essentially of transformations between inertial frames, but in 
view of the goal, we should avoid using the relativity principle in the 
derivation of such transformations.

We assume, that there is {\em at least one} inertial frame in which light 
behaves isotropically. We call it the priviledged frame $\Sigma$ and denote 
space and time coordinates in this frame by the capital letters: $(X,Y,Z,T)$. 
In $\Sigma$, clocks are synchronised with Einstein's 
procedure \cite{eins:05a}.
We consider also an other system $S$ moving with uniform 
velocity $v<c$ along 
the $X$-axis in the positive direction. In $S$, the coordinates are written 
with 
lower case letters $(x,y,z,t)$.
Under rather general assumptions, symmetry conditions on the two systems, 
the assumption that the two-way velocity of light is $c$ and furthermore
that the time 
dilation factor has its relativistic value, one can derive
the following transformation (see \cite{mase:77a}, \cite{sell:94a}):
\begin{eqnarray}
x & = & \frac{1}{\sqrt{1-\beta^{2}}}\left(X-vT\right) \nonumber \\
y & = & Y \nonumber \\
z & = & Z \label{eq:stp} \\
t & = & s\left(X-vT\right) + \sqrt{1-\beta^{2}}\;\; T\;\;, \nonumber
\end{eqnarray}
where $\beta=v/c$. The parameter $s$, which determines the synchronisation 
in the $S$ frame remains unknown. Einstein's synchronisation 
in $S$ involves: 
\mbox{ $s=-v/(c^{2}\sqrt{1-\beta^{2}})$} and (\ref{eq:stp}) becomes a Lorentz 
boost. For a general $s$, the inverse one-way velocity of light is given
by:
\begin{equation}
\frac{1}{c_{\rightarrow}(\Theta)} = \frac{1}{c} + \left(\frac{\beta}{c} +
s\sqrt{1-\beta^{2}}\right)\cos\Theta\;\;,
\label{eq:lexo}
\end{equation}
where $\Theta$ is the angle between the $x$-axis and the light ray in $S$.
$c_{\rightarrow}(\Theta)$ is in general dependent on the direction. 
A simple case is $s=0$. 
This means from (\ref{eq:stp}), that at $T=0$ of $\Sigma$ we set all clocks 
of $S$ 
at $t=0$ (external synchronisation), or that we synchronise the clocks by means 
of light rays with velocity $c_{\rightarrow}(\Theta)=c/(1+\beta\cos\Theta)$
(internal synchronisation).
It should be stressed that, unlike to the parameters of lenght contraction 
and time dilation, {\em the parameter s cannot be tested}, but its 
value must be 
assigned in accordance with the synchronisation chosen in the experimental 
setup. For a recent and comprehensive discussion of this subject, see 
\cite{vest:93a}. A striking consequence of (\ref{eq:lexo}) is that the negative 
result 
of the Michelson-Morley experiment does not rule out an ether. Only an ether 
with galilean transformations is excluded, because the galilean 
transformations  
do not lead to an invariant two-way velocity of light in a moving system.

\section{One-way velocity of light on the rim of a disk}

We now study synchronisation on the rim 
of a rotating disk. This problem has numerous applications, such as the 
Sagnac effect
and the synchronisation of atomic clocks around the earth.
Let consider a disk rotating anticlockwise with constant angular 
velocity $\omega$ 
respective to the priviledged frame $\Sigma$ (for simplicity), around the 
$Z$-axis. Suppose  
light is constrained to move on the rim of the disk and that ideal 
clocks are 
put on that rim. 
Because of the {\em Clock Hypothesis} ideal clocks and rods on the rim 
behave exactly in the same 
way as in the instantaneously coinciding tangential inertial frame. Since 
it is 
obvious, at least 
tangentially to the rim, that the velocity of light is the same in the 
coinciding tangential inertial frame as on the corresponding 
part of the rim (the 
gravitational field is perpendicular to the rim), {\em we must choose the 
same 
synchronisation in these two frames}. 
Let us now construct a metric on the whole disk, using the transformations:
\begin{equation}
T  =  t;\;\;
X  =  r\cos(\phi +\omega t);\;\;
Y  =  r\sin(\phi +\omega t);\;\;  
Z  =  z
\label{eq:haldol}
\end{equation}
The coordinates $(t,r,\phi,z)$ give a possible description of the physical 
events for an observer at rest on the disk. In particular, the 
coordinate $t$
is measured with a clock that runs $(1-\omega^{2}r^{2}/c^{2})^{-1/2}$ faster
than a clock at rest in $\Sigma$, so that we have the right to write the 
first equation of
(\ref{eq:haldol}). From (\ref{eq:haldol}) we obtain the line element:
\begin{equation}
ds^{2}=\left(1-\frac{\omega^{2}r^{2}}{c^{2}}\right)\left(cdt\right)^{2}
-2\frac{\omega r^{2}}{c} d\phi\left(cdt\right)
-dr^{2}-r^{2}d\phi^{2}-dz^{2}
\label{eq:reypnol}
\end{equation}
Let remark that the metric is not static.
As is well known, the spatial part of the metric is not only given by the 
space-space coefficients of the four dimensional metric, but by:
\begin{equation}
dl^{2}=\left(-g_{\alpha\beta}+\frac{g_{0\alpha}g_{0\beta}}{g_{00}}\right)
dx^{\alpha}dx^{\beta}
=dr^{2}+dz^{2}+\frac{r^{2}d\phi^{2}}{1-\frac{\omega^{2}r^{2}}{c^{2}}},
\label{eq:mescaline}
\end{equation}
where $0$ is the time index, and $\alpha,\beta$ represent the space indices 
and can take the values $1,2,3$. The right-hand-side of (\ref{eq:mescaline}), 
with $dz=0$,
is the standard result, showing that the spatial part of the metric is not 
flat on the rotating disk. 

We now use an Einstein's synchronisation on the disk. Generally, if we send 
a light signal from point $A$ with coordinates $x^{\alpha},\;\alpha=1,2,3$ 
to an infinitesimally near point $B$ with coordinates 
$x^{\alpha}+dx^{\alpha},\;\alpha=1,2,3$ and back, the coordinate time 
difference $dt_{1}$ ($dt_{2}$) for the ``there'' (back) trip is 
obtained by 
solving the equation $ds^{2}=0$. We obtain:
\begin{equation}  
dt_{1,2}  =  \frac{1}{cg_{00}}\left[\mp g_{0\alpha}dx^{\alpha}+\sqrt{\left(
g_{0\alpha}g_{0\beta}-g_{\alpha\beta}g_{00}\right)
dx^{\alpha}dx^{\beta}}\right]
 =  \pm\frac{\omega r^{2} d\phi}{c^{2}\left(
1-\frac{\omega^{2}r^{2}}{c^{2}}\right)}
    +\frac{dl_{AB}}{c \sqrt{1-\frac{\omega^{2}r^{2}}{c^{2}}}}
\label{eq:datura}
\end{equation}
By definition the time $t_{A}$ at $A$ which is synchronous with the 
arrival time in $t_{B}$ in $B$ is the midtime of departure and 
arrival at
$A$. So, two Einstein-synchronous events are not coordinate-time- 
synchronous 
and have a difference $\Delta t$ such that: 
\begin{equation}
t_{B}=t_{A}+\Delta t = t_{A} 
-\frac{1}{c}\frac{g_{0\alpha}dx^{\alpha}}{g_{00}}=t_{A}+\frac{\omega 
r^{2}d\phi}{c^{2}\left(1-\frac{\omega^{2}r^{2}}{c^{2}}\right)}
\label{eq:amphete}
\end{equation}
If we generalise this procedure, not only in an infinitesimal domain, but 
along 
a curve, we obtain that generally it is path dependant, because 
$\Delta t$ is not a total differential. As consequence, time 
can not be defined globally on the rim with Einstein's procedure. This is 
what 
in fact happens on earth: if one synchronises atomic clocks all around the 
earth with Einstein's procedure and comes back to the point of departure 
after a 
whole round trip, a time lag will result. This means that a clock is not 
synchronisable with itself, which is clearly absurd. 
Hence the physicist Ashby from 
Boulder has said: 
``Thus one discards Einstein's synchronisation in the rotating 
frame''\cite{ashb:93a}. 
On the other hand, we can easily define a global time since the coordinate 
time 
$t$ is already global.
Remembering that the coordinate time $t$ is measured with clocks that run 
faster than clocks at rest in $\Sigma$, we define a global 
time $t'=\sqrt{1-\omega^{2}r^{2}/c^{2}}\;t$. 
The one-way velocity of light on the rim is 
now given by:
\begin{equation}
c_{\rightarrow}(\pm)=\frac{dl_{AB}}{dt'_{1,2}}=
\frac{dl_{AB}}{dt_{1,2} \sqrt{1-\frac{\omega^{2}r^{2}}{c^{2}}}}=
\frac{c}{1\pm\frac{\omega r}{c}}\;\;,
\label{eq:volubilis}
\end{equation}
where the last step comes from (\ref{eq:mescaline}) and (\ref{eq:datura}) 
with $dr=dz=0$ and $+$ ($-$) stands for the anticlockwise (clockwise) 
propagation of light.
It means that the velocity of light in the tangential inertial frame is also 
equal to:
$c_{\rightarrow}(\pm)=c/(1\pm\beta)$, with $\beta=\omega r/c$
, corresponding to a parameter $s=0$ of (\ref{eq:lexo}) and an angle
$\Theta$ of $0$ ($\pi$).

\section{Conclusion}
In inertial systems, the synchronisation of clocks is conventional.
When extended to accelerated systems, the synchronisation is 
no longer conventional \cite{sell:96a}. A consistent theory must involve a 
one-way velocity of light which depends on the frame of reference.
Transformations (\ref{eq:lexo}) with values of the synchronisation 
parameter other than $s=0$ 
lead, from a physical point of view, to an inadequate description of 
time on the rim of a rotating disk.

\section{Acknowledgements}

I want to thank the Physics Departement of Bari University for hospitality.

\end{document}